\documentclass{article}
\usepackage{spconf,amsmath,epsfig}
\usepackage[utf8]{inputenc}
\usepackage{times}
\usepackage{epsfig}
\usepackage{graphicx}
\usepackage{amsmath}
\usepackage{amssymb}
\usepackage{arydshln}
\usepackage{enumitem}

\usepackage{multirow}
\usepackage{booktabs}

\usepackage[pagebackref=true,breaklinks=true,colorlinks,bookmarks=false]{hyperref}

\begin{document}\sloppy

\def\x{{\mathbf x}}
\def\L{{\cal L}}

\title{Cross-modal Embeddings for Video and Audio Retrieval}
%

\name{Didac Surís\(^1\), Amanda Duarte\(^2\), Amaia Salvador\(^1\), Jordi Torres\(^2\) and Xavier Giró-i-Nieto\(^1\)}
\address{\(^1\)Universitat Politècnica de Catalunya (UPC) \\ 
\(^2\)Barcelona Supercomputing Center (BSC-CNS)}


\maketitle

\begin{abstract}

The increasing amount of online videos brings several opportunities for training self-supervised neural networks. The creation of large scale datasets of videos such as the YouTube-8M allows us to deal with this large amount of data in manageable way. In this work, we find new ways of exploiting this dataset by taking advantage of the multi-modal information it provides. By means of a neural network, we are able to create links between audio and visual documents, by projecting them into a common region of the feature space, obtaining joint audio-visual embeddings. These links are used to retrieve audio samples that fit well to a given silent video, and also to retrieve images that match a given a query audio. The results in terms of Recall@K obtained over a subset of YouTube-8M videos show the potential of this unsupervised approach for cross-modal feature learning. We train embeddings for both scales and assess their quality in a retrieval problem, formulated as using the feature extracted from one modality to retrieve the most similar videos based on the features computed in the other modality.

\end{abstract}



\section{Introduction}
Videos have become the next frontier in artificial intelligence. The rich semantics contained in them make them a challenging data type posing several challenges in both perceptual, reasoning or even computational level.
Mimicking the learning process and knowledge extraction that humans develop from our visual and audio perception remains an open research question, and video contain all this information in a format manageable for science and research.

Videos are used in this work for two main reasons. Firstly, they naturally integrate both visual and audio data, providing a weak labeling of one modality with respect to the other.
Secondly, the high volume of both visual and audio data allows training machine learning algorithms whose models are governed by a high amount of parameters.
The huge scale video archives available online and the increasing number of video cameras that constantly monitor our world, offer more data than computation power available to process them. 


The popularization of deep neural networks among the computer vision and audio communities has defined a common framework boosting multimodal research. 
Tasks like video sonorization, speaker impersonation or self-supervised feature learning have exploited the opportunities offered by artificial neurons to project images, text and audio in a feature space where bridges across modalities can be built.


This work exploits the relation between the visual and audio contents in a video clip to learn a joint embedding space with deep neural networks. Two multilayer perceptrons (MLPs), one for visual features and a second one for audio features, are trained to be mapped into the same cross-modal representation. We adopt a self-supervised approach, as we exploit the unsupervised correspondence between the audio and visual tracks in any video clip.

\begin{figure}
   	\includegraphics[width=\columnwidth]{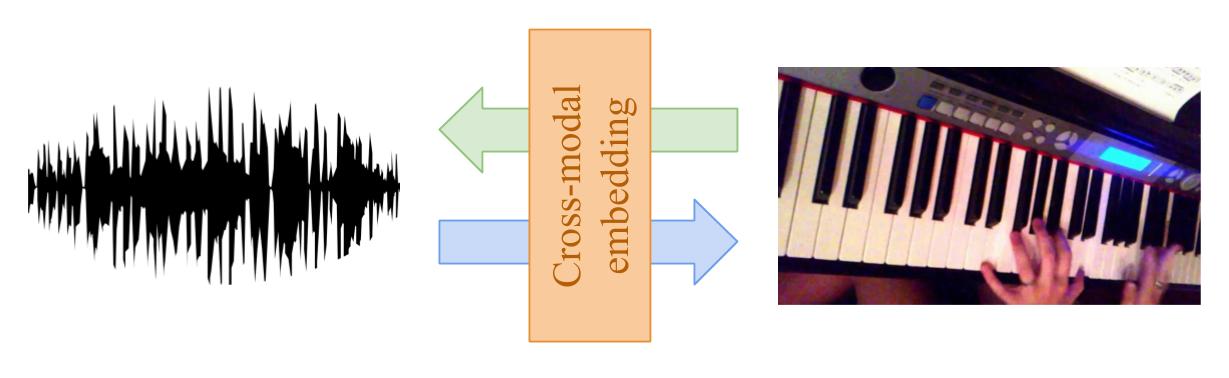}
    \caption{A learned cross-modal embedding allows retrieving video from audio, and vice versa.}
    \label{fig:intro}
\end{figure}

We propose a joint audiovisual space to address a retrieval task formulating a query from any of the two modalities. As depicted in Figure \ref{fig:intro}, whether a video or an audio clip can be used as a query to search its matching pair in a large collection of videos. For example, an animated GIF could be sonorized by finding an adequate audio track, or an audio recording illustrated with a related video.

In this paper, we present a simple yet very effective model for retrieving documents with a fast and light search. We do not address an exact alignment between the two modalities that would require a much higher computation effort.

The paper is structured as follows. Section \ref{sec:related_work} introduces the related work on learned audiovisual embeddings with neural networks.  Section \ref{sec:architecture} presents the architecture of our model and Section \ref{sec:training} how it was trained. Experiments are reported in Section \ref{sec:experiments} and final conclusions drawn in Section \ref{sec:conclusions}. The source code and trained model used in this paper is publicly available from \url{https://github.com/surisdi/youtube-8m}.

\section{Related work}\label{sec:related_work}


In the past years, the relationship between the audio and the visual content in videos has been researched in several contexts. Overall, conventional approaches can be divided into four categories according to the task: generation, classification, matching and retrieval.

As online music streaming and video sharing websites have become increasingly popular, some research has been done on the relationship between music and album covers \cite{brochu2003sound, mayer2011analysing, libeks2011you, chao2011tunesensor} and also on music and videos (instead of just images) as the visual modality \cite{schindler2015audio, wu2016bridging, acar2014understanding, gillet2007correlation} to explore the multimodal information present in both types of data. 

A recent study \cite{li2003multimedia} also explored the cross-modal relations between the two modalities but using images with people talking and speech. It is done through Canonical Correlation Analysis (CCA) and cross-modal factor analysis. Also applying CCA, \cite{zhang2007cross} uses visual and sound features and common subspace features for aiding clustering in image-audio datasets. In a work presented by \cite{ngiam2011multimodal}, the key idea was to use greedy layer-wise training with Restricted Boltzmann Machines (RBMs) between vision and sound.


The present work is focused on using the information present in each modality to create a joint embedding space to perform cross-modal retrieval. This idea has been exploited especially using text and image joint embeddings \cite{WangLL15, KirosSZ14, salvador2017learning}, but also between other kinds of data, for example creating a visual-semantic embedding \cite{41869} or using synchronous data to learn discriminative representations shared across vision, sound and text \cite{aytar2017see}.

However, joint representations between the images (frames) of a video and its audio have yet to be fully exploited, being \cite{DBLP:journals/corr/HongIY17} the work that most has explored this option up to the knowledge of the authors. In their paper, they seek for a joint embedding space but only using music videos to obtain the closest and farthest video given a query video, only based on either image or audio.

The main idea of the current work is borrowed from \cite{salvador2017learning}, which is the baseline to understand our approach. There, the authors create a joint embedding space for recipes and their images. They can then use it to retrieve recipes from any food image, looking to the recipe that has the closest embedding. Apart from the retrieval results, they also perform other experiments, such as studying the localized unit activations, or doing arithmetics with the images.


\section{Architecture}\label{sec:architecture}

\begin{figure*}[t]
  \begin{center}
   	\includegraphics[width=\textwidth]{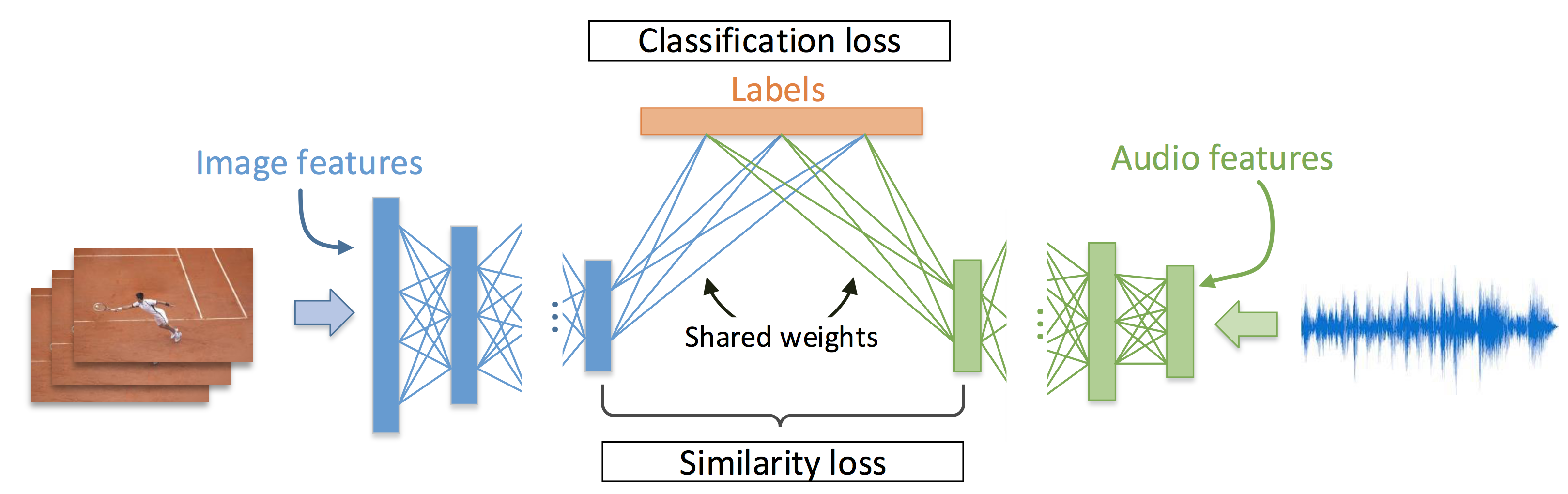}
    \caption{Schematic of the used architecture.}
    \vspace*{-\baselineskip}
    \label{fig:schematic}
  \end{center}
\end{figure*}

In this section we present the architecture for our joint embedding model, which is depicted in the Figure \ref{fig:schematic}. 

As inputs, we have the vector of features representing the images of the video and the vector of features representing the audio. These features are already precomputed and provided in the YouTube-8M dataset \cite{youtube8m}. In particular, we use the \textit{video-level} features, which represent the whole video clip with two vectors: one for the audio and another one for the video. These feature representations are the result of an average pooling of the local audio features computed over windows of one second, and local visual features computed over frames sampled at 1 Hz.


The main objective of the system is to transform the two different features (image and audio, separately) to other features laying in a \textit{joint space}. This means that for the same video, ideally the image features and the audio features will be transformed to the same joint features, in the same space. We will call these new features \textbf{embeddings}, and will represent them with $\Phi^{i}$, for the image embeddings, and $\Phi^{a}$, for the audio embeddings.

The idea of the joint space is to represent the \textit{concept} of the video, not just the image or the audio, but a generalization of it. As a consequence, videos with similar concepts will have closer embeddings and videos with different concepts will have embeddings further apart in the joint space. For example, the representation of a tennis match video will be close to the one of a football match, but not to the one of a maths lesson.

Thus, we use a set of fully connected layers of different sizes, stacked one after the other, going from the original features to the embeddings. The audio and the image network are completely separated. These fully connected layers perform a non-linear transformation on the input features, mapping them to the embeddings, being the parameters of this non-linear mapping learned in the optimization process.

After that, a classification from the two embeddings is done, also using a fully connected layer from them to the different classes, using a sigmoid as activation function. We will get more insight on this step in section \ref{sec:training}.

The number of hidden layers is not necessarily fixed, as well as the number of neurons per layer, since we experimented with different configurations. 
Each hidden layer uses ReLu as activation function, and all the weights in each layer are regularized using L2 norm.

\section{Training}\label{sec:training}

In this section we present the used losses as well as their meaning and intuition.

\subsection{Similarity Loss}

The objective of this work is to get the two embeddings of the same video to be as close as possible (ideally, the same), while keeping embeddings from different videos as far as possible. 

Formally, we are given a video $v_k$, represented by the audio and visual features $v_k=\{i_k, a_k\}$ ($i_k$ represents the image features and $a_k$ the audio features of $v_k$). The objective is to maximize the similarity between $\Phi^i_k$, the embedding obtained by transformations on $i_k$, and $\Phi^a_k$, the embedding obtained by transformations on $a_k$.

At the same time, however, we have to prevent embeddings from different videos to be ``close'' in the joint space. In other words, we want them to have low similarity. However, the objective is not to force them to be \textit{opposite} to each other. Instead of forcing them to have similarity equal to zero, we allow a margin of similarity small enough to force the embeddings to be clearly not in the same place in in the joint space. We call this margin $\alpha$.

During the training, both positive and negative pairs are used, being the positive pairs the ones for which $i_k$ and $a_k$ correspond to the same video $v_k$, and the negative pairs the ones for which $i_{k1}$ and $a_{k2}$ do not correspond to the same video, this is, $k1\neq k2$. The proportion of negative samples is $p_{\text{negative}}$.

For the negative pairs, we selected random pairs that did not have any common label, in order to help the network to learn how to distinguish different videos in the embedding space.
The notion of ``similarity'' or ``closeness'' is mathematically translated into a cosine similarity between the embeddings, being the cosine similarity defined as:
\vspace*{-\baselineskip}
\begin{equation}
similarity = \cos(x,z) = \frac{\sum\limits_{k=1}^N{x_k z_k}}{\sqrt{\sum\limits_k^N{x_k^2}}\sqrt{\sum\limits_i^N{z_k^2}}}
\end{equation}

for any pair of real-valued vectors $x$ and $z$.

From this reasoning we get to the first and most important loss:

\begin{equation}
\begin{split}
L_{cos}((\Phi^a, &\Phi^i), y) =\\
&=\left\{ \begin{array}{lcc}
             1-\cos(\Phi^a, \Phi^i), &\text{if  } & y = 1 \\
             \max(0, \cos(\Phi^a, \Phi^i) - \alpha), &\text{if} & y = -1 \\
             \end{array}
   \right.
   \end{split}
\end{equation}
\label{eq:cosine_loss}

where $y=1$ denotes positive sampling, and $y=-1$ denotes negative sampling.

\subsection{Classification Regularization}

Inspired by the work presented in \cite{salvador2017learning}, we provide additional information to our system by incorporating the video labels (classes) provided by the YouTube-8M dataset. 
This information is added as a regularization term that seeks to solve the high-level classification problem, both from the audio and from the video embeddings, sharing the weights between the two branches. The key idea here is to have the classification weights from the embeddings to the labels shared by the two modalities. 

This loss is optimized together with the previously explained similarity loss, serving as a regularization term. Basically, the system learns to classify the audio and the images of a video (separately) into different classes or labels provided by the dataset. We limit its effect by using a regularization parameter $\lambda$.

To incorporate the previously explained regularization to the joint embedding, we use a single fully connected layer, as shown in Figure \ref{fig:schematic}. 
Formally, we can obtain the label probabilities as $p^i = \text{softmax}(W\Phi^i)$ and $p^a = \text{softmax}(W\Phi^a)$, where $W$ represents the learned weights, which are shared between the two branches. 
The \textit{softmax} activation is used in order to obtain probabilities at the output. The objective is to make $p^i$ as similar as possible to $c^i$, and $p^a$ as similar as possible to $c^a$, where $c^i$ and $c^a$ are the category labels for the video represented by the image features and the audio features, respectively. For positive pairs, $c^i$ and $c^a$ are the same.

The loss function used for the classification is the well known cross entropy loss:
\begin{equation}
L(x, z)=-\sum\limits_k x_k \log(z_k)
\end{equation}

Thus, the classification loss is:
\begin{equation}
L_{class}(p^i, p^a, c^i, c^a)=-\sum\limits_k (p^i_k \log(c^i_k) + (p^a_k \log(c^a_k) )
\end{equation}
\label{eq:classification_loss}


Finally, the loss function to be optimized is:

\begin{equation}
L = L_{cos} + \lambda L_{class}
\end{equation}
\label{eq:general_loss}
\vspace*{-\baselineskip}
\subsection{Parameters and Implementation Details}

For our experiments we used the following parameters: \begin{itemize}[noitemsep]
\item Batch size of $1024$.
\item We saw that starting with $\lambda$ different than zero led to a bad embedding similarity because the classification accuracy was preferred. Thus, we began the training with $\lambda=0$ and set it to $0.02$ at step number 10,000.
\item Margin $\alpha = 0.2$.
\item Percentage of negative samples  $p_{\text{negative}}$ = 0.6.
\item 4 hidden layers in each network branch, the number of neurons per layer being, from features to embedding, 2000, 2000, 700, 700 in the image branch, and  450, 450, 200, 200 in the audio branch.
\item Dimensionality of the feature vector = 250.
\item We trained a single epoch.
\end{itemize}


The simulation was programmed using Tensorflow \cite{abadi2016tensorflow}, having as a baseline the code provided by the YouTube-8M challenge authors\footnote{\url{https://www.kaggle.com/c/youtube8m}}.

\section{Results}\label{sec:experiments}

\subsection{Dataset}

The experiments presented in this section were developed over a subset of 6,000 video clips from the YouTube-8M dataset \cite{youtube8m}. This dataset does \emph{not} contain the raw video files, but their representations as precomputed features, both from audio and video.
Audio features were computed using the method explained in \cite{7952132} over audio windows of 1 second, while visual features were computed over frames sampled at 1 Hz with the Inception model provided in TensorFlow \cite{abadi2016tensorflow}.



The dataset provides \textit{video-level} features, which represent all the video using a single vector (one for audio and another for visual information), and thus does not maintain temporal information; and also provides \textit{frame-level} features, which consist on a single vector representing each second of audio, and a single vector representing each frame of the video, sampled at 1 frame per second. 


The main goal of this dataset is to provide enough data to reach state of the art results in video classification. Nevertheless, such a huge dataset also permits approaching other tasks related to videos and cross-modal tasks, such as the one we approach in this paper. For this work, and as a baseline, we only use the \textit{video-level} features. 

\subsection{Quantitative Performance Evaluation}

We divide our results in two different categories: quantitative (numeric) results and qualitative results. 

To obtain the quantitative results we use the Recall@k metric. We define Recall@k as the recall rate at top K for all the retrieval experiments, this is, the percentage of all the queries where the corresponding video is retrieved in the top K, hence higher is better.

The experiments are performed with different dimension of the feature vector. The Table \ref{table:recall_at_k_av} shows the results of recall from audio to video. In other words, from the audio embedding of a video, how many times we retrieve the embedding corresponding to the images of that same video. Table \ref{table:recall_at_k_va} shows the recall from video to audio.

To have a reference, the random guess result would be $k/\text{Number of elements}$. The obtained results show a very clear correspondence between the embeddings coming from the audio features and the ones coming from the video features. It is also interesting to notice that the results from audio to video and from video to audio are very similar, because the system has been trained bidirectionally.

\begin{table}
		\caption{Evaluation of Recall from audio to video}
		\label{table:recall_at_k_av}
		\begin{tabular}{c|c c c}
          Number of elements & Recall@1 & Recall@5 & Recall@10 \\
          \hline
          256 & 21.5\% & 52.0\% & 63.1\% \\
          512 & 15.2\% & 39.5\% & 52.0\% \\
          1024 & 9.8\% & 30.4\% & 39.6\% \\
		\end{tabular}
\end{table}

\begin{table}
		\caption{Evaluation of Recall from video to audio}
		\label{table:recall_at_k_va}
		\begin{tabular}{c|c c c}
          Number of elements & Recall@1 & Recall@5 & Recall@10 \\
          \hline
          256 & 22.3\% & 51.7\% & 64.4\% \\
          512 & 14.7\% & 38.0\% & 51.5\% \\
          1024 & 10.2\% & 29.1\% & 40.3\% \\
		\end{tabular}
\end{table}

\subsection{Qualitative Performance Evaluation}

In addition to the objective results, we performed some insightful qualitative experiments. They consisted on generating the embeddings of both the audio and the video for a list of 6,000 different videos. Then, we randomly chose a video, and from its image embedding, we retrieved the video with the closest audio embedding, and the other way around (from one video's audio we retrieved the video with the closest image embedding). If the closest embedding corresponded to the same video, we took the second one in the ordered list.

\begin{figure*}[!ht]
  \begin{center}
   	\includegraphics[width=\textwidth]{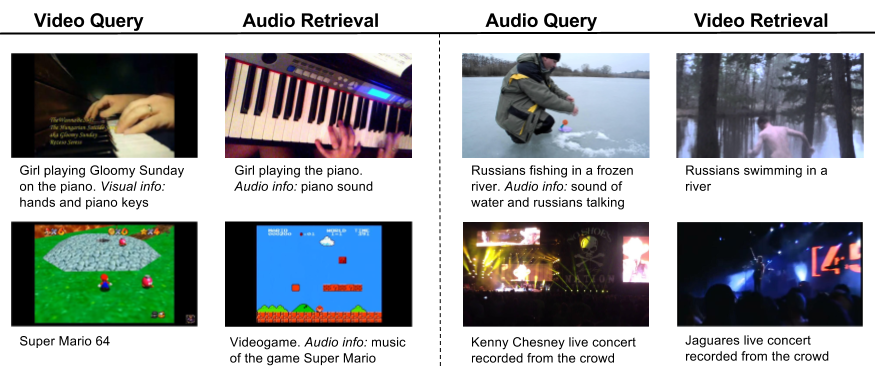}
    \vspace*{-\baselineskip}
    \caption{Qualitative results. On the left we show the results obtained when we gave a video as a query. On the right, the results are based on an audio as a query.}
    \label{fig:experiments}
  \end{center}
\end{figure*}

The Figure \ref{fig:experiments} shows some experiments. On the left, we can see the results given a video query and getting the closest audio; and on the right the input query is an audio. Examples depicting the real videos and audio are available online \footnote{\url{https://goo.gl/NAcJah}}. It shows both the results when going from image to audio, and when going from audio to image. Four different random examples are shown in each case. For each result and each query, we also show their YouTube-8M labels, for completeness. 

The results show that when starting from the image features of a video, the retrieved audio represents a very accurate fit for those images. Subjectively, there are non negligible cases where the retrieved audio actually fits better the video than the original one, for example when the original video has some artificially introduced music, or in cases where there is some background commentator explaining the video in a foreign (unknown) language. This analysis can also be done similarly the other way around, this is, with the \textit{audio colorization} approach, providing images for a given audio.

\section{Conclusions and Future Work}\label{sec:conclusions}
We presented an effective method to retrieve audio samples that fit correctly to a given (muted) video. The qualitative results show that the already existing online videos, due to its variety, represent a very good source of audio for new videos, even in the case of only retrieving from a small subset of this large amount of videos. Due to the existing difficulty to create new audio from scratch, we believe that a retrieval approach is the path to follow in order to give audio to videos.

The range of possibilities to extend the presented work is excitingly broad. The first idea would be to make use of the YouTube-8M dataset variety and information. The temporal information provided by the individual image and audio features is not used in the current work. The most promising future work implies using this temporal information to match audio and images, making use of the implicit synchronization the audio and the images of a video have, without needing any supervised control. Thus, the next step in our research is introducing a recurrent neural network, which will allow us to create more accurate representations of the video, and also retrieve different audio samples for each image, creating a fully synchronized system.

Also, it would be very interesting to study the behavior of the system depending on the class of the input. Observing the dataset, it is clear that not all the classes have the same degree of correspondence between audio and image, as for example some videos have artificially (posterior) added music, which is not related at all to the images.

In short, we believe the YouTube-8M dataset allows for promising research in the future in the field of video sonorization and audio retrieval, for it having a huge amount of samples, and for it capturing multi-modal information in a highly compact way.

\section{Acknowledgements}\label{sec:acknowledgements}
This work was partially supported by the Spanish Ministry of Economy and Competitivity and the European Regional Development Fund (ERDF) under contract TEC2016-75976-R. Amanda Duarte was funded by the mobility grant of the Severo Ochoa Program at Barcelona Supercomputing Center (BSC-CNS).

\bibliographystyle{IEEEbib}
\bibliography{icme2018template}

\end{document}